\documentclass[letterpaper, 10 pt, conference,romanappendices]{ieeeconf}  

\IEEEoverridecommandlockouts                              

\usepackage{url,color,amsmath,graphicx,enumerate}

\usepackage{amsmath} 
\usepackage{amssymb}  
\usepackage{dsfont}
\usepackage{bm}
\usepackage{pgfplots}

\usepackage{filecontents}

\newtheorem{theorem}{Theorem}
\newtheorem{lemma}{Lemma}
\newtheorem{corollary}{Corollary}

\title{\LARGE \bf
A Two Stage Stochastic Mechanism for Selling Random Power
}
\author{Nathan Dahlin and Rahul Jain
\thanks{Nathan Dahlin and Rahul Jain are with the Department of Electrical Engineering, University of Southern California, Los Angeles, CA 90089
        {\tt\small dahlin@usc.edu; rahul.jain@usc.edu}}%
}

\begin{document}

\maketitle
\thispagestyle{empty}
\pagestyle{empty}

\begin{abstract}
We present a two stage auction mechanism that renewable generators (or aggregators) could use to allocate renewable energy among LSEs. The auction is conducted day-ahead. LSEs submit bids specifying their valuation per unit, as well as their real-time fulfillment costs in case of shortfall in generation. We present an allocation rule and a de-allocation rule that maximizes expected social welfare. Since the LSEs are strategic and may not report their private valuations and costs truthfully, we design a two-part payment, one made in Stage 1, before renewable energy generation level $W$ is realized, and another determined later to be paid as compensation to those LSEs that have to be ``de-allocated" in case of a shortfall. We proposes a two-stage Stochastic VCG mechanism which we prove is incentive compatible in expectation (expected payoff maximizing bidders will bid truthfully), individually rational in expectation (expected payoff of all participants is non-negative) and is also efficient. To the best of our knowledge, this is the first such two-stage mechanism for selling random goods.
\end{abstract}

\section{INTRODUCTION}\label{sec:intro}


Remarkable strides have been made in meeting the aggressive renewable energy portofolio standards worldwide. For example, California has gone from under 5\% of energy from renewable sources in 2010 to nearly 25\% today and expected to reach 33\% in 2020. In fact, the state has mandated 100\% of energy to be derived from renewable sources by 2045. Other states and countries are following suit \cite{cal100percent}. This presents immense technological and economic challenges.

Renewable energy sources such as wind and solar are inherently variable, meaning that efforts to bridge such gaps must address the challenge of efficiently mitigating the impact of uncertainty in generation. For some time, renewable resources have been considered more akin to negative loads than firm block generating units, and their growth has been spurred by feed-in tariffs \cite{abbad2010electricity}, \cite{demeo2005wind}. In order to maintain grid stability, it is the responsibility of an independent system operator (ISO) to procure adequate reserves to compensate for potential shortfalls in generation. While feasible at relatively low levels of usage, at higher levels of penetration such arrangements hamper the net benefits of renewable energy \cite{bitar2012selling}. Therefore, prevailing approaches to renewable integration have shifted from so called supply-push-type mechanisms \cite{abbad2010electricity} to those which place more of the burden on the generators themselves as well as their end users. 

This shift has manifested in a couple of ways, first via what can be categorized as technological regulations. For example in Spain, both new and existing generators are required to equip themselves with fault-ride-through capability. Secondly, economic regulations have evolved such that generators are more directly exposed to the risks associated with their variability. Again turning to the Spanish example, mandatory hourly forecasting requirements were implemented along with penalties of 10\% estimated system cost for deviations of over 20\%. 

This example typifies two-settlement systems: renewable generators are forced to participate in conventional energy markets with \textit{ex-post} monetary penalties for deviations from \textit{ex-ante} contracts. The latter are settled in day-ahead (DA) markets, while the former are determined in real-time (RT) spot markets. Formally, the determination of committed quantities and allocations can be captured in the framework of optimization problems \cite{morales2013integrating}.

Adopting the view of wind or solar energy as a random good, much work has focused on how the manner in which generators bring their product to market can affect social welfare. In \cite{tan1993interruptible} and \cite{bitar2012selling}, given the probability distribution of power generation scenarios, contracts of the form $(\rho_{k},p_k)$, signifying a price of $p_k$ per unit of the contract, which will be fulfilled with probability $\rho_k$, must be designed. Thus, Load serving entitities (LSEs) are given an opportunity on how much risk due to insufficient generation they need to take on, given the variable-reliability options presented by the supplier. Such a market is shown in \cite{bitar2012selling} to operate more efficiently than a firm-electricity market, which incurs additional costs including the provisioning of reserves. In \cite{bitar2012bringing} the problem is to set a price $p$ and offered generation quantity $C$ in order to maximize expected profit, rather than social welfare. Still it is shown that the optimal expected shortfall is nondecreasing in $p$ and $C$, demonstrating the need to curtail offerings to reduce necessary reserve capacities. 

The studies above assume that the distribution used in determining optimal offerings is reported truthfully. \cite{tang2015market} investigates the aggregator's task of selecting a subset of available generators to maximize a given objective (e.g. maximize expected generation) as well as the ISO's problem of pricing wind energy, given a set of available generators, allowing for strategic behavior by generators in reporting their generation distributions. A stochastic VCG mechanism, as well as a commitment with penalty type mechanism are proposed which elicit truthful distribution reporting, i.e., incentive compatibility, satisfy generator individual rationality (or voluntary participation) constraints and achieve efficient outcomes. 

In this paper, we examine the situation wherein a renewable generator, or an aggregator allocates generated electricity among a set of LSEs via an auction mechanism. We design a two stage auction. In Stage 1 (day ahead), the LSEs make bids that specify their value $v$ for each unit. Furthermore, they also specify their cost $c$ of real-time fulfillment (e.g., from the spot market) in case there is a shortfall and the generator cannot meet the commitment it made in Stage 1. In Stage 2, random generation level $W$ is realized. In case there is a shortfall over the commitment already made in Stage 1, the auctioneer ``de-allocates" some of the LSEs but pays them a compensation that depends on the real-time fulfillment costs $c$. Since the LSEs are strategic and need not report their values $v$ and costs $c$ truthfully, we have to devise the allocation, de-allocation and payment rules such that it leaves no incentive for the LSEs to not be truthful. Furthermore, we would also like such a mechanism to be efficient in the sense of maximizing the expected social welfare even in the absence of knowledge about the true valuations and costs of the LSEs.

We note that the problem posed here is a two stage auction mechanism for allocation of a random good with two part payments. Literature on multi-stage and dynamic mechanisms is sparse since they are usually regarded as rather difficult problems. The reader may refer to \cite{mas1995microeconomic} for the standard game theoretic terminology that we use throughout the paper.

\section{Preliminaries}\label{sec:prelim}

\subsection{The Setting}
Consider a renewable energy generator with random integer valued generation $W\in [0,\overline{w}]$ where $\overline{w}\in\mathbb{Z}^+$ gives the maximum generation amount. Let $p = (p_0,\dots,p_{\overline{w}})$ give the probability mass function for the generator's output. We will consider a two stage setting. In stage 1, the generator conducts an auction to sell the random renewable energy generation in which $N$ load serving entities (LSEs) participate and determines an allocation. In stage 2, $W$ is realized, and say $W=k$ with probability $p_k$. This is known to both the generator and the LSEs. This may result in a deficit over the allocation in stage 1, in which case some of the LSEs do not receive any power but may incur some cost to fulfill demand from spot market. Denote the value of receiving a unit by $v_i$ and the cost of unfulfilled demand (after allocation in stage 1)  by $c_i$ for LSE $i$. Further denote $\sigma=(\sigma_1,\dots\sigma_N)$, which is LSE $i$'s private information. 


The generator thus, faces a two-stage stochastic optimization problem. In stage 1, it decides out of the various bidding LSEs, which ones to commit to (the allocation). And once renewable energy generation $W$ has been realized, if there is shortfall, in stage 2 it decides which LSEs will receive energy units, and which one's will get de-allocated. Let us denote stage 1 decision by $x=x(\sigma,p)$ and stage 2 decision by $z=z(x,w)$. $x_i=1$ denotes allocation to LSE $i$ in stage 1, and zero otherwise. $z_i=1$ if  LSE $i$ is deallocated in stage 2, and zero otherwise. 

Thus given a first stage decision $x$ and realized generation level $W=w$, the generator's second stage decision problem can be formulated as 
\begin{alignat}{2}\label{secondstage}\min_{z}&\quad\sum_{i=1}^N(v_i+c_i)z_i\\
&\quad z_i\leq x_i\quad\forall\,i \nonumber\\
&\quad \sum_{i=1}^Nz_i=\left(\sum_{i=1}^Nx_i-w\right)_+\nonumber
\end{alignat}
where $x_+ = \max(x,0)$. Let $Q(x,w;\sigma)$ denote the minimum cost achievable in (\ref{secondstage}). 

Now define the \textit{social welfare} (SW) achieved when allocation $x$ is chosen and generation level $k$ is realized as 
\begin{equation}\label{SWxk}SW(x,w;\sigma)=\sum_{i=1}^Nv_ix_i - Q(x,w;\sigma)\end{equation}


The deallocation $z$ depends on shortfall in generation $W$ which is random. Thus, in stage 1 the generator maximizes expected social welfare to determine allocation $x$:
\begin{equation}\label{ESW}\max_{x}\quad\mathbb{E}[SW(x,W;\sigma)]\end{equation}
Note that (\ref{ESW}) may also be written
\begin{equation}\label{optprob_orig}\max_{x}\quad\sum_{i=1}^Nv_ix_i - \mathbb{E}[Q(x,w;\sigma)]\end{equation}

\subsection{Mechanism Design}

The generator now holds an auction in which it asks the various LSEs to submit bids. LSE $i$ submits a bid $\hat{\sigma}_i=(\hat{v}_i,\hat{c}_i)$ which may be different from its private information $\sigma_i=(v_i,c_i)$.  The generator will use the bids $\hat{\sigma}=(\hat{\sigma}_1,\dots\hat{\sigma}_N)$ to make allocation (and deallocation) decisions. But the allocation can only be expected to be efficient, i.e., social welfare maximizing if the bidders truthfully report $\sigma_i=(v_i,c_i)$. Let $\gamma_i=v_i+c_i$ and similarly $\hat{\gamma}_i=\hat{v}_i+\hat{c}_i$. But bidders are strategic and unless provided proper incentives need not be truthful. Thus, we would like to design an auction mechanism that aligns incentives of LSEs such that they indeed are truthful. 

To specify such a mechanism, we need to specify the allocation rule (based on the bids), the deallocation rule (based on the bids and the generation realization $w$) and the payments to be made. Let $\mathcal{I}\subseteq N$ be the subset of LSEs selected to receive a potential unit of energy. Those LSEs which are not selected in $\mathcal{I}$ leave the auction with zero payoff. Each selected LSE $i$ makes a payment $t^d_i(\hat{\sigma})$ to the generator, prior to realization of $W$. Then in the RT market, after realization of $W$, say $W=w$, a final subset $\mathcal{I}_w\subseteq\mathcal{I}$ is selected to receive any available units. Additionally each LSE $i\in\mathcal{I}$ receives a payment $t^r_i(\hat{\sigma},w)$ from the generator. Both $t^r_i$ and $t^d_i$ are allowed to take on negative values, indicating a payment in the reverse direction specified. Together these selection and payment schemes define a direct revelation mechanism $\Gamma = (\mathcal{I},\mathcal{I}_w,t^d_i,t^r_i)$.

In order to specify the desired properties of such an auction mechanism it is necessary to first define the payoffs each LSE will receive. Therefore we define the payoff of LSE $i$, given generator decisions $(x,z)$, generation realization $w$ and bids $\hat{\sigma}$ as
\begin{equation}\pi_i(x,z,w;\sigma) = \hat{v}_ix_i-(\hat{v}_i+\hat{c}_i)z_i-t^d_i+t^r_i\end{equation}

The first goal in designing our auction mechanism will be to ensure LSE participation. We assume that this can be accomplished if no LSE can expect to receive a negative payoff, and say that a mechanism achieves \textit{individual rationality (IR) in expectation} when:
\begin{equation}\label{IRinEx}\mathbb{E}_W[\pi_i(x,z,W;\sigma)] \geq 0\quad \forall\,i \end{equation}

(\ref{IRinEx}) states that each LSE can expect to receive a nonnegative payoff when participating in the mechanism. Given that the LSEs choose to participate in the mechanism, it is further desired that they bid truthfully. This will occur if it is in their interest to do so, and we say that a mechanism achieves \textit{incentive compatibility (IC) in expectation} when:
\begin{equation}\label{ICinEx}\mathbb{E}_W[\pi_i(x,z,W;(\sigma_i,\hat{\sigma}_{-i})]\geq \mathbb{E}_W[\pi_i(x,z,W;(\hat{\sigma}_i,\hat{\sigma}_{-i})]\end{equation}
for all $i$ and $\hat{\sigma}_{-i}$, where $\hat{\sigma}_{-i}=(\hat{\sigma}_1,\dots,\hat{\sigma}_{i-1},\hat{\sigma}_{i+1},\dots,\hat{\sigma}_N)$ gives the bid profile aside from $\hat{\sigma}_i$. (\ref{ICinEx}) states that for each LSE $i$, regardless of the bids of the other LSEs, bidding truthfully yields at least as high an expected value as any other strategy. 

Finally, given that LSEs choose to participate truthfully it is desired that the auction mechanism makes LSE selections which maximize the expected social welfare in (\ref{ESW}). We say that an auction mechanism is \textit{efficient} if it selects an allocation such that (\ref{ESW}) is maximized. 

Again, the generator needs to know the valuations and costs of the LSEs in order to select the best set of LSEs, in the sense of social welfare maximization. As the LSEs are strategic, they may not truthfully reveal their valuations and costs. Thus the goal of this work is to design an auction mechanism which is individually rational and incentive compatible in expectation, as well as efficient. As the mechanism to be designed is one-sided, we do not attempt to achieve budget balance. 

\section{The Generator's Problem}\label{sec:generator}

Given that our mechanism will be given in the form $\Gamma=(\mathcal{I},\mathcal{I}_k,t^d_i,t^r_i)$, we now give a reformulation of the generator's problem. In stage 1, it decides which LSEs to allocate to, and in stage 2, with knowledge of the generation realization, it decides which LSEs of those allocated in stage 1, need to be deselected (and compensated for) when there is a generation shortfall. Defining $x$ as previously we have that 
\begin{equation}\label{selectscheme}\mathcal{I} = \{i\,:\,x_i=1\}\end{equation}
The number of LSEs selected will be denoted $n=|\mathcal{I}|$. Let $X^{-i}$ denote the set of allocations for which $x_i=0$. Note that functions of $x$ can be considered equivalently as functions of $\mathcal{I}$, so that occasionally we will interchange them as arguments. For example $Q(x,w;\sigma)\equiv Q(\mathcal{I},w;\sigma)$. 

Upon realization of the value of $W$, the generator makes a second stage selection of the final subset of LSEs $\mathcal{I}_w\subseteq \mathcal{I}$ to receive the available units, where the subscript $w$ reflects that $W$ has assumed the value $w\in\{0,1,\dots,\overline{w}\}$. This leaves a subset $\overline{\mathcal{I}}_w=\mathcal{I}\setminus \mathcal{I}_w$ of deselected LSEs. Therefore defining $z$ as in the previous discussion we have 
\begin{equation}\label{secondstage_select}\mathcal{I}_w=\{i\,:\,z_i=0\}\end{equation}

Let $\mathcal{N}$ denote the power set of $\{0,1,\dots,N\}$. Let $\hat{\gamma}_{i}:=(\hat{v}_{i}+\hat{c}_{i})$. Then the generator's two stage problem can be rewritten as
\begin{equation}\label{optprob}
\begin{split}
&\max_{\mathcal{I}\in\mathcal{N}}\quad\sum_{i\in\mathcal{I}}\hat{v}_i-\sum_{w=0}^{|\mathcal{I}|-1}p_w\min_{\overline{\mathcal{I}}_w\subseteq\mathcal{I}}\left\{\sum_{i\in \overline{\mathcal{I}}_w}\hat{\gamma}_i\,:\,|\overline{\mathcal{I}}_w|=|\mathcal{I}|-w\right\}
\end{split}
\end{equation}
Note that assuming truthful bids, the inner minimization problem in (\ref{optprob}) is equivalent to (\ref{secondstage}), and the outer maximization problem is equivalent (\ref{optprob_orig}).


Therefore, in solving (\ref{optprob_orig}), the generator determines its selection $\mathcal{I}$. Given selection $\mathcal{I}$ with $n=|\mathcal{I}|$, only when generation level $w<n$ will the minimum cost in (\ref{secondstage}) be positive. Otherwise, each LSE the generator made a commitment to in stage 1 will receive a unit, yielding $Q(x,w;\hat{\sigma})=0$. When the realization $w<n$, the generator will solve (\ref{secondstage}), to determine which LSE are to be deselected with decision $z$ and associated $\overline{\mathcal{I}}_w$.


Note that the selection of a particular $\mathcal{I}$ induces a ranking on the selected LSEs. For simplicity assume that the values of $\hat{\gamma}$ are unique. Suppose that $n=|\mathcal{I}|$ LSEs have been selected in the first stage, and $w<n$ units are generated. $\overline{\mathcal{I}}_w$ thus, will include the $n-w$ LSEs selected in $\mathcal{I}$ with the lowest $\hat{\gamma}_i$. This implies that for each LSE $i\in\mathcal{I}$ there is a maximum level of generation at which they will not receive a unit. For example the LSE with rank 1 will not receive a unit if zero units are generated. The LSE with rank 2 will not receive a unit if one or fewer units are generated and so on. If this maximum level is denoted $w_i$ for $i\in\mathcal{I}$, then let $r_i(\mathcal{I}):= r_i :=  w_i+1$ denote the \textit{rank} of LSE $i$ for $i\in\mathcal{I}$. 

We use the notation $(\cdot)$ to indicate indexing with reference to the selection $\mathcal{I}$. For example the reported valuation of the LSE with rank 1 under $\mathcal{I}$ is denoted $\hat{v}_{(1)}$. Thus, when $|\mathcal{I}|=n$
\begin{equation}\hat{\gamma}_{(1)}> \hat{\gamma}_{(2)}>\dots>\hat{\gamma}_{(n-1)}> \hat{\gamma}_{(n)}\end{equation}
For $j\notin\mathcal{I}$ let $r_j:=\overline{w}+1$. We will say that LSE $(1)$ occupies the highest rank and LSE $(n)$ occupies the lowest rank, given $\mathcal{I}$. 

\section{Stochastic VCG Mechanism for Selling Random Power}\label{sec:mechanism}

We introduce a stochastic VCG mechanism for selling random power. To specify our mechanism $\Gamma$ we need to specify selection schemes $\mathcal{I}$ and $\mathcal{I}_w$ and payment schemes $t^d_i$ and $t^r_i$. The solutions to (\ref{optprob_orig}) and (\ref{secondstage}) give $\mathcal{I}$ and $\mathcal{I}_w$. We now give the payment schemes $t^r_i$ and $t^d_i$. 

First, define LSE $i$'s utility under allocation $x$ given production level $w$ as
\begin{equation}\label{utility}u_{i}(\mathcal{I},\hat{\sigma},w):=u_{i}(\hat{\sigma},w)=x_i(v_{i}-\gamma_{i}\mathds{1}_{\{w<r_i\}})\end{equation}
Note that the valuation and cost terms in (\ref{utility}) are the true values for LSE $i$, not the reported ones. LSE $i$'s payoff is given as 
\begin{equation}\pi_{i}(\hat{\sigma},w)=u_{i}(\hat{\sigma},w)-t_{i}(\hat{\sigma},w)\end{equation}
where $t_{i}(\hat{\sigma},w) = t^d_i(\hat{\sigma}) -t^r_i(\hat{\sigma},w)$. Taking expectation over all generation scenarios gives
\begin{align}\mathbb{E}[\pi_{i}(\hat{\sigma},W)] &= \mathbb{E}[u_{i}(\hat{\sigma},W)] - \mathbb{E}[t_{i}(\hat{\sigma},W)]\\
&= \mathbb{E}[x_i\left(v_{i} - \gamma_{i}\mathds{1}_{\{w<r_i\}}\right)] - \mathbb{E}[t_{i}(\hat{\sigma},W)]\\
&= x_i\left(v_{i} - \gamma_{i}\mathbb{E}[\mathds{1}_{\{w<r_i\}}]\right) - \mathbb{E}[t_{i}(\hat{\sigma},W)]\\
&= x_i\left(v_{i} - \gamma_{i}\sum_{w=0}^{r_i-1}p_w\right) - \mathbb{E}[t_{i}(\hat{\sigma},W)]
\end{align}

Now, to specify the payment rules $t^r_i$ and $t^d_i$ we need a couple of additional definitions. First, fixing a first stage decision $\mathcal{I}$ and LSE $(i)\in\mathcal{I}$, for $j\notin\mathcal{I}$, denote
\begin{equation}\begin{split}\label{thetadef}\theta^{i}_j &:= \bigg(\hat{v}_j - \hat{\gamma}_jp_0\\
&-\sum_{w=1}^{i-1}p_w\min(\hat{\gamma}_{(w)},\hat{\gamma}_{j})-\sum_{w=i}^{n-1}p_w\min(\hat{\gamma}_{(w+1)},\hat{\gamma}_{j})\bigg)\end{split}\end{equation}
and
\begin{equation}\begin{split}\label{argmax}j(i) &:= \underset{j\notin\mathcal{I}}{\arg\max}\,\theta^{i}_j\end{split}\end{equation}
where we assume that $j(i)$ is unique. 

$\theta^{i}_j$ represents the contribution that a particular LSE $j\notin\mathcal{I}$ would make to the expected social welfare, if it were selected along with LSEs $\mathcal{I}\backslash\{(i)\}$. If $\theta^i_j>0$ for at least one LSE $j\notin\mathcal{I}$, then $j(i)$ is the LSE that is selected if LSE $i$ is disregarded, forming selection $\mathcal{I}\backslash\{(i)\}\cup\{j(i)\}$. Denote $\theta^{i}_{j(i)}:= \overline{\theta}^i$, and similarly $\hat{v}_{j(i)}:=\overline{v}^i$ and $\hat{\gamma}_{j(i)}:=\overline{\gamma}^i$. 

The motivation for $j(i)$ is as follows. Having fixed $\mathcal{I}$ we will want to determine the externality that LSE $(i)$ imposes on the generator and other LSEs. This externality will be a function of the best selection possible given that $x_{(i)}=0$, i.e., the best solution which disregards LSE $(i)$. Denote this selection $\mathcal{I}^{-(i)}$. As will be shown, $\mathcal{I}^{-(i)}$ will select all other LSEs in $\mathcal{I}$, i.e., $\mathcal{I}\backslash\{(i)\}$ as well as up to one additional LSE $j\notin\mathcal{I}$, depending upon whether $\overline{\theta}^{i}>0$. This LSE is $j(i)$.

As described in section III, the selection $\mathcal{I}^{-(i)}$ will induce a ranking on the included LSEs. The notation $(\cdot)^{-(i)}$ will be used to refer to this ranking. In particular, $r_{j(i)}(\mathcal{I}^{-(i)}):= \overline{r}^i$ gives the rank of LSE $j(i)$ amongst LSEs $\mathcal{I}^{-(i)}$. 

The payments $t^d_{(i)}$ and $t^r_{(i)}$ for LSE ${(i)}\in\mathcal{I}$ will depend upon $\overline{\theta}^{i}$ and $\overline{r}^{i}$ and are specified in Table I. A positive $t^d_{(i)}$ indicates a payment from LSE ${(i)}$ to the generator in the DA market, and a positive $t^r_{(i)}$ indicates a payment from the generator to LSE ${(i)}$ in the RT market. In either case negative values indicate transfers in the opposite direction. If LSE $j\notin\mathcal{I}$ then $t^d_j=t^r_j=0$. 

\begin{table}[htbp]
\begin{center}
\caption{Payment function for LSE ${(i)}\in\mathcal{I}$}
\begin{tabular}{|c|c|l|}
\hline  Cases& $t^d_{(i)}(\hat{\sigma})$ & $\quad\quad\quad\quad\quad\quad t^r_{(i)}(\hat{\sigma},w)$ \\
\hline 1. $\begin{array}{c}\overline{\theta}^{i}\leq 0\end{array}$ & 0 & $\begin{array}{ll}
0&\text{$0\leq w\leq i-1$}\\
-\hat{\gamma}_{(w+1)}&\text{$i\leq w\leq n-1$}\\
0&\text{$w\geq n$}\end{array}$ \\
\hline 2. $\begin{array}{c} \overline{\theta}^{i}>0\\  
\overline{r}^{i}>i\end{array}$ & $\overline{v}^{i}$ & $\begin{array}{ll}
\overline{\gamma}^{i}&\text{$0\leq w\leq i-1$}\\
\overline{\gamma}^{i}-\hat{\gamma}_{(w+1)}&\text{$i\leq w\leq \overline{r}^{i}-1$}\\
0&\text{$\overline{r}^{i}\leq w$}
\end{array} $ \\
\hline 3. $\begin{array}{c}\overline{\theta}^{i}>0\\  \overline{r}^{i}\leq i\end{array}$ & $\overline{v}^{i}$ & $\begin{array}{ll}
\overline{\gamma}^{i}&\text{$0\leq w\leq \overline{r}^{i}-1$}\\
\hat{\gamma}_{(w)}&\text{$\overline{r}^{i}\leq w\leq i-1$}\\
0&\text{$w\geq i$}
\end{array} $ \\
\hline 
\end{tabular}
\end{center}
\end{table}

\begin{figure}
\begin{tikzpicture}
\begin{axis}[
    axis x line=center, 
    axis y line=middle, 
    xtick={0,1,2,3,4,5,6,7,8},
    xticklabels={0,1,2,3,4,5,6,7,8},
    ytick={-4,-2.75,-1.75},
    yticklabels={$-\hat{\gamma}_{(4)}$,$-\hat{\gamma}_{(5)}$,$-\hat{\gamma}_{(6)}$},
    domain=0:9,
    range = -6:2,
    xlabel=$w$,
    ylabel=$t^{r}_{(3)}$,
    ymin = -6,
    ymax = 3,
    xmin = 0,
    xmax = 8,
    width=8cm,
    height=4cm
]
\addplot+[ycomb] plot table[x expr=\coordindex, y index=1]{datafile1.dat};
\end{axis}
\end{tikzpicture}
\caption{Case 1 RT transfer for $i=3$ and $n=6$. }
\end{figure}
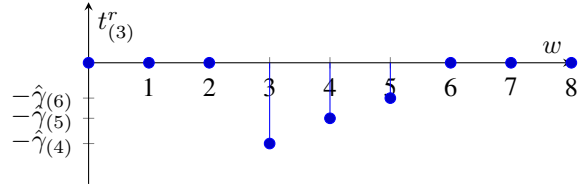


\textit{Case 1}: $(\overline{\theta}^{i}\leq 0)$ Since $\overline{\theta}^{i}\leq 0$, adding any LSE $j\notin\mathcal{I}$ to the selection $\mathcal{I}\backslash\{(i)\}$ will not increase the expected social welfare. Thus LSE $(i)$ does not cause an externality in the first stage and $t^d_{(i)} = 0$. In the second stage, when $w\leq i-1$, LSE $(i)$ is deselected and $t^{r}_{(i)}=0$. If $i\leq w\leq n-1$ then LSE $(i)$ receives a unit, but there is a shortfall and LSE $(i)$ pays $\hat{\gamma}_{(w+1)}$, the sum of value lost and cost incurred by LSE $(w+1)$, the ``first'' deselected LSE. When $w\geq n$, LSE $(i)$ and all other selected LSEs receive a unit and $t^r_{(i)}=0$. 

\begin{figure}
\begin{tikzpicture}
\begin{axis}[
    axis x line=center, 
    axis y line=middle, 
    xtick={0,1,2,3,4,5,6,7,8},
    xticklabels={0,1,2,3,4,5,6,7,8},
    ytick={2,-1.25},
    yticklabels={$\overline{\gamma}^i$,$\overline{\gamma}^i-\hat{\gamma}_{(4)}$},
    domain=0:8,
    range = -6:2,
    xlabel=$w$,
    ylabel=$t^{r}_{(3)}$,
    ymin = -2,
    ymax = 3,
    xmin = 0,
    xmax = 7,
    width=8cm,
    height=4cm
]
\addplot+[ycomb] plot table[x expr=\coordindex, y index=1]{datafile2.dat};
\end{axis}
\end{tikzpicture}
\caption{Case 2 RT transfer for $i=3$, $n=6$ and $\overline{r}^3=4$. }
\end{figure}
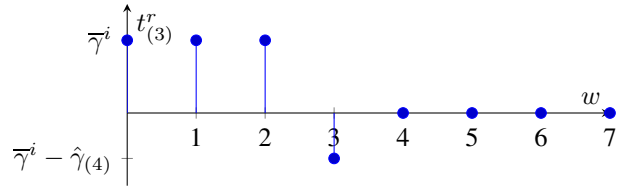

\textit{Case 2}: $(\overline{\theta}^{i}>0,\overline{r}^{i}>i\,)$ $\overline{\theta}^{i}>0$ indicates that if LSE $(i)$ were not present then some other LSE $j\notin\mathcal{I}$ could be added to the selection $\mathcal{I}\backslash\{(i)\}$ to increase the expected SW. In particular LSE $j(i)$ could be added to maximize this increase, so in the DA market LSE $i$ pays $t^d_{(i)}=\overline{v}^{i}$, the value term that LSE $j(i)$ would have added to the expected SW. 

If $w\leq i-1$, then LSE $(i)$ does not receive a unit, but received compensation $\overline{\gamma}^i$. If $i\leq w\leq \overline{r}^i-1$ then LSE $(i)$ receives a unit, but makes a payment of $t^r_{(i)}=\overline{\gamma}^i-\hat{\gamma}_{(w+1)}$. To see that this difference is nonpositive observe that given $w\in[i, \overline{r}^i-1]$, since $\hat{\gamma}_{(j)}$ is increasing in $j$, $\hat{\gamma}_{(w+1)}\geq \hat{\gamma}_{(\overline{r}^i)}\geq \overline{\gamma}^i$. The last inequality here follows from the assumption $\overline{r}^i\leq i-1$, which implies that LSE $(i+1)$ has rank $i$ in selection $\mathcal{I}\backslash\{(i)\}\cup j(i)$, i.e. $r_{(i+1)}(\mathcal{I}^{-i})=i>\overline{r}^i$.   

If $w\geq \overline{r}^i$, then $t^r_{(i)}=0$. This is because while LSE $(i)$ is omitted, they are replaced by LSE $j(i)$, so that all LSEs with rank higher than $\overline{r}^i$ in selection $\mathcal{I}$ will have the same rank in selection $\mathcal{I}^{-(i)}$. Therefore the presence of LSE $(i)$ in the RT market does not affect whether or not they experience shortfall, i.e. LSE $(i)$ imposes no externality on them. 

\begin{figure}
\begin{tikzpicture}
\begin{axis}[
    axis x line=center, 
    axis y line=middle, 
    xtick={0,1,2,3,4,5,6,7,8},
    xticklabels={0,1,2,3,4,5,6,7,8},
    ytick={3,2},
    yticklabels={$\overline{\gamma}^i$,$\hat{\gamma}_{(2)}$},
    domain=0:8,
    range = -6:2,
    xlabel=$w$,
    ylabel=$t^{r}_{(3)}$,
    ymin = -2,
    ymax = 4,
    xmin = 0,
    xmax = 7,
    width=8cm,
    height=4cm
]
\addplot+[ycomb] plot table[x expr=\coordindex, y index=1]{datafile3.dat};
\end{axis}
\end{tikzpicture}
\caption{Case 3 RT transfer for $i=3$, $n=6$ and $\overline{r}^3=2$. }
\end{figure}

\textit{Case 3}: $(\overline{\theta}^i>0,\overline{r}^i\leq i\,)$ The scheme and explanation for $t^d_{(i)}$ in this case is the same as in the previous case. 

When $0\leq w\leq \overline{r}^i-1<i$, LSE $(i)$ receives no unit, but they are compensated with a payment $t^r_{(i)}=\overline{\gamma}^i$. Alternatively this may be interpreted as LSE $i$ receiving a credit of $\overline{\gamma}^{i}$, as their presence prevents the selection of LSE $j(i)$, and therefore the loss of $\overline{\gamma}^{i}$ when $w<\overline{r}^i$.

If $\overline{r}^i\leq w\leq i-1$, then LSE $(i)$ again receives no unit and is compensated with $t^r_{(i)} = \hat{\gamma}_{(w)}$. Again this can be interpreted in terms of hypothetical loss. Since $w\geq \overline{r}^i$, LSE $j(i)$ would receive a unit. Since $\overline{r}^i\leq i$, LSE $j(i)$ would force LSEs $(\overline{r}^i)$ through $(i-1)$ to one higher rank, i.e., $r_{(k)}(\mathcal{I}^{-i})=k+1$ for $\overline{r}^i\leq k\leq i-1$. Therefore, if $\overline{r}^i\leq w\leq i-1$, LSE $(w)$ would be the ``first'' LSE to be deselected under $\mathcal{I}^{-i}$, and so LSE $(i)$ imposes a positive externality of $\hat{\gamma}_{(w)}$. 

Finally, if $w\geq i$ then LSE $(i)$ receives a unit and $t^r_{(i)}=0$. In this case, no RT transfer occurs between LSE $(i)$ and the generator because as in Case 2, LSE $(i)$ is omitted and then replaced in $\mathcal{I}^{-i}$, with $\overline{r}^i<i$. Therefore, the rank of LSEs $(i+1)$ through $(n)$ does not change, and the presence of LSE $(i)$ imposes no externality on them.

Note that for the case where $\overline{r}^i=i$, the payments given by the last two rows of Table 1 agree, i.e., either scheme may be used. The following example illustrates this point. 

\textit{Example 1}: Let $\hat{v}_1=3$, $\hat{v}_2=2$ and $\hat{v}_3 = 13/32$, and $\hat{\gamma}_1 = 2$, $\hat{\gamma}_2=1$ and $\hat{\gamma}_3 = 1/2$. Assume that the generation distribution is given by $p = (1/2,1/4,1/8,1/8)$. Then $\mathcal{I}=\{1,2\}$, and LSE 1 has rank 1 and LSE 2 has rank 2. To see why this selection is optimal, observe that if LSE 3 were also selected, then since $\hat{\gamma}_3<\hat{\gamma}_2<\hat{\gamma}_1$, in the scenarios where $w\in\{0,1,2\}$, they would be deselected. That is $\mathcal{I}_{0}=\emptyset$, $\mathcal{I}_1=\{1\}$, $\mathcal{I}_2 = \{1,2\}$ and $\mathcal{I}_3=\{1,2,3\}$. Therefore, their expected contribution to social welfare would be given by 
\begin{equation}\hat{v}_{3}-\hat{\gamma}_3(p_0+p_1+p_2)=13/32 - (1/2)(7/8) <0,\end{equation}
while LSE 2 contributes
\begin{equation}\hat{v}_2 - \hat{\gamma}_2(p_0+p_1) = 2 - 1(3/4) >0,\end{equation}
and LSE 1 contributes
\begin{equation}\hat{v}_1 - \hat{\gamma}_1p_0 = 3 - (2)(1/2) > 0.\end{equation}
However if either LSE 1 or 2 were not selected, and LSE 3 were selected, then it would achieve rank 2 and make a contribution of
\begin{equation}\overline{\theta}^1=\overline{\theta}^2=\hat{v}_{3}-\hat{\gamma}_3(p_0+p_1)=13/32 - (1/2)(3/4) >0.\end{equation}
to the expected social welfare. 
Thus LSE 1 and LSE 2 fall under case 2 and 3, respectively, of Table I, and their DA payments are
\begin{equation}t^d_1=t^d_2 = 13/32.\end{equation}
For LSE 1 $t^r_1(\hat{\sigma},0)=1/2$, $t^r_1(\hat{\sigma},1) = 1/2-1=-1/2$ and $t^r_1(\hat{\sigma},w)=0$ for $w\geq2$. For LSE 2, $t^r_2(0)=t^r_2(1) = 1/2$, and $t^r_2(w)=0$ for $w\geq 2$. $\hfill\blacksquare$

We call the mechanism consisting of selection scheme (\ref{selectscheme}), deselection scheme (\ref{secondstage_select}) and payments listed in Table 1 for $i\in\mathcal{I}$ and $t^d_i := 0$, $t^r_i := 0$ for $i\notin\mathcal{I}$ a Stochastic VCG Mechanism for Random goods ({\tt SVCG-RANDOM}).

\begin{theorem}\label{thm:main}
The {\tt SVCG-RANDOM} mechanism is individually rational and incentive compatible in expectation, as well as efficient. 
\end{theorem}

This appears to be the first two stage mechanism for random goods such as renewable energy. While we focus on LSEs wanting single units, it can be extended to multi-unit domain and bids. IR and IC properties are achieved in expectation ex ante. Ex post IR and IC are unlikely to be achievable. 

\section{Proof of Theorem \ref{thm:main}}

Subsection B of this section presents a proof of the claimed mechanism properties in each of the three cases listed in Table I. The proof relies on a lemma, from subsection A, concerning the optimal selection that the generator makes when $x_{(i)}=0$, denoted $\mathcal{I}^{-(i)}$. 

\subsection{The Selection $\mathcal{I}^{-(i)}$}

In order to show that the proposed selection and payment schemes constitute a mechanism achieving the properties stated in Theorem 1, an explicit form of $\mathcal{I}^{-(i)}$ for $(i)\in\mathcal{I}$ is needed. Below, Lemma 2 enumerates the possible forms of $\mathcal{I}^{-(i)}$, as well as conditions for when each form applies. Lemma 1 is necessary in the proof of Lemma 2. Proofs of both may be found in the Appendix.  

\begin{lemma}\label{thm:lemma1}
For any LSE $j\notin\mathcal{I}$ with $n=|\mathcal{I}|$
\begin{equation}\hat{v}_j-\hat{\gamma}_jp_0\leq \sum_{w=1}^{n}\min(\hat{\gamma}_j,\hat{\gamma}_{(w)})p_w\leq\hat{\gamma}_j \sum_{w=1}^np_w\end{equation}
\end{lemma}
\begin{lemma}\label{thm:lemma2}
For $(i)\in\mathcal{I}$, the optimal selection given $x_{(i)}=0$, $\mathcal{I}^{-(i)}$, is 
\begin{equation}
\mathcal{I}^{-(i}) = 
\begin{cases}
\mathcal{I}\backslash\{(i)\}&\text{if $\overline{\theta}^i\leq 0$}\\
\mathcal{I}\backslash\{(i)\}\cup\{j(i)\}&\text{if $\overline{\theta}^i>0$}
\end{cases}
\end{equation}
\end{lemma}

\subsection{Proof of Theorem \ref{thm:main}}

Let $x^{-i}$ denote the first stage allocation corresponding to the solution of (\ref{optprob}) when $x_{i}=0$. Let $r^{-i}_j$ give the ranking of LSE $j$ under selection $x^{-i}$, and $V^{-i}$ denote the maximum expected SW achieved when $x_i=0$, given bids $\hat{\sigma}$. 

For some LSE $i$, fix $\hat{\sigma}_{-i}$, the bids of the other LSEs, and assume that LSE $i$ bids truthfully, so that the bid vector is $\hat{\sigma} = (\sigma_i,\hat{\sigma}_{-i})$. Also assume that given this $\hat{\sigma}$, the generator makes selection $\mathcal{I}$ with $n=|\mathcal{I}|$, and that LSE $i\in\mathcal{I}$ with rank $i$, so that in this context $(i)=i$. Let $V^*$ be the optimal expected SW associated with selection $\mathcal{I}$. 

\textit{Case 1}: $(\overline{\theta}^{i}\leq 0)$ Since LSE $i\in\mathcal{I}$, $x_{i}=1$ and 
\begin{equation}\begin{split}&\mathbb{E}[\pi_{i}(\mathcal{I},W;\hat{\sigma})] = v_{i}-\gamma_{i}\mathbb{E}[\mathds{1}_{\{W<i\}}] - t^d_{i}+\mathbb{E}[t^r_{i}(W)]\\
&=v_{i}-\gamma_{i}\sum_{w=0}^{i-1}p_w - \sum_{w=i}^{n-1}\hat{\gamma}_{(w+1)}p_w\\
&=v_{i}-\gamma_{i}\sum_{w=0}^{i-1}p_w + \left(\sum_{i\neq j}x_j\left(\hat{v}_j-\hat{\gamma}_j\sum_{w=0}^{r_j-1}p_w\right)\right)\\
&\label{lastline}\qquad\quad-\left(\sum_{i\neq j}x_j\left(\hat{v}_j-\hat{\gamma}_j\sum_{w=0}^{r_j-1}p_w\right)+\sum_{w=i}^{n-1}\hat{\gamma}_{(w+1)}p_w\right)
\end{split}\end{equation}
\begin{equation}\begin{split}
&=v_{i}-\gamma_{i}\sum_{w=0}^{r_{i}-1}p_w + \left(\sum_{i\neq j}x_j\left(\hat{v}_j-\hat{\gamma}_j\sum_{w=0}^{r_j-1}p_w\right)\right)\\
&-\left(\sum_{\substack{j\leq n\\j\neq i}}\left(\hat{v}_{(j)}-\hat{\gamma}_{(j)}\sum_{w=0}^{j-1}p_w\right)+\sum_{w=i}^{n-1}\hat{\gamma}_{(w+1)}p_w\right)
\end{split}\end{equation}
\begin{equation}\label{exppayoff_v_i}\begin{split}
=&x_{i}\left(v_{i}-\gamma_{i}\sum_{w=0}^{r_{i}-1}p_w\right) + \sum_{i\neq j}x_j\left(\hat{v}_j-\hat{\gamma}_j\sum_{w=0}^{r_j-1}p_w\right)\\
&\qquad-V^{-i}
\end{split}\end{equation}
\begin{equation}
=V^*-V^{-i}\geq 0,
\end{equation}
where the second to last equality holds due to Lemma 2, and the last holds due to $\hat{\sigma}=(\sigma_i,\hat{\sigma}_{-i})$. Note that the first line of (\ref{exppayoff_v_i}) changes with $\hat{\sigma}_i$ only via $r$ and $x$. If LSE $i$ bids truthfully, then the generator will choose the $x$ and $r$ which maximize these terms. Given some $\hat{\sigma}_i\neq \sigma_i$, the generator will make a different selection, which can only decrease these terms. Therefore truthful bidding is a weakly dominant strategy which also ensures a nonnegative expected payoff for LSE $(i)$. Thus for Case 1 and LSE $i\in\mathcal{I}$ the mechanism is IC and IR in expectation. 

\textit{Case 2}: $(\overline{\theta}^i< 0, \overline{r}^i>i)$

\begin{equation}\begin{split}&\mathbb{E}[\pi_{i}(\mathcal{I},W;\hat{\sigma})] = v_{i}-\gamma_{i}\mathbb{E}[\mathds{1}_{\{W<i\}}] - t^d_{i}+\mathbb{E}[t^r_{i}(W)]\\
&= v_{i}-\gamma_{i}\sum_{w=0}^{i-1}p_w - \overline{v}^{i}+\overline{\gamma}^i\sum_{w=0}^{i-1}p_w+\sum_{w=i}^{\overline{r}^i-1}(\overline{\gamma}^i-\hat{\gamma}_{(w+1)})p_w\\
\end{split}\end{equation}
\begin{equation}\begin{split}
&=x_i\left(v_{i}-\gamma_{i}\sum_{w=0}^{r_i-1}p_w\right)+\sum_{j\neq i}x_j\left(\hat{v}_j-\hat{\gamma}_j\sum_{w=0}^{r_j-1}p_w\right)\\
&-\Bigg[\sum_{\substack{j\leq n\\j\neq i}}\left(\hat{v}_{(j)}-\hat{\gamma}_{(j)}\sum_{w=0}^{r_i-1}p_w\right)+\sum_{w=i}^{\overline{r}^i-1}\hat{\gamma}_{(w+1)}p_w\\
&\qquad\qquad\qquad\qquad\qquad\qquad+\overline{v}^i-\overline{\gamma}^i\sum_{w=0}^{\overline{r}^i-1}p_w\Bigg]
\end{split}
\end{equation}
\begin{equation}\begin{split}
&=x_i\left(v_{i}-\gamma_{i}\sum_{w=0}^{r_i-1}p_w\right)+\sum_{j\neq i}x_j\left(\hat{v}_j-\hat{\gamma}_j\sum_{w=0}^{r_j-1}p_w\right)\\
&-\Bigg[\sum_{\substack{j\leq n\\j\neq i}}\left(\hat{v}_{(j)}-\hat{\gamma}_{(j)}\sum_{w=0}^{r_i-1}p_w\right)+\sum_{w=i}^{\overline{r}^i-1}\hat{\gamma}_{(w+1)}p_w+\overline{\theta}^i\Bigg]\end{split}
\end{equation}
\begin{equation}\label{exppayoff_v_i2}\begin{split}
&=x_i\left(v_{i}-\gamma_{i}\sum_{w=0}^{r_i-1}p_w\right)+\sum_{j\neq i}x_j\left(\hat{v}_j-\hat{\gamma}_j\sum_{w=0}^{r_j-1}p_w\right)-V^{-i}\\
&=V^*-V^{-i}\geq 0.\end{split}
\end{equation}

Again the second to last equality holds due to Lemma 2. By the same argument as for Case 1, LSE $i$'s expected payoff is maximized when they bid truthfully, and this maximized expected payoff is nonnegative. Thus for Case 3 and LSE $i\in\mathcal{I}$ the mechanism is IC and IR in expectation. 

\textit{Case 3}: $(\overline{\theta}^i< 0, \overline{r}^i\leq i)$

\begin{equation}\begin{split}&\mathbb{E}[\pi_{i}(\mathcal{I},W;\hat{\sigma})] = v_{i}-\gamma_{i}\mathbb{E}[\mathds{1}_{\{W<i\}}] - t^d_{i}+\mathbb{E}[t^r_{i}(W)]\\
&= v_{i}-\gamma_{i}\sum_{w=0}^{i-1}p_w - \overline{v}^{i}+\overline{\gamma}^i\sum_{w=0}^{\overline{r}^i-1}p_w+\sum_{w=\overline{r}^i}^{i-1}\hat{\gamma}_{(w)}p_w\\
\end{split}\end{equation}
\begin{equation}\begin{split}
&=x_i\left(v_{i}-\gamma_{i}\sum_{w=0}^{r_i-1}p_w\right)+\sum_{j\neq i}x_j\left(\hat{v}_j-\hat{\gamma}_j\sum_{w=0}^{r_j-1}p_w\right)\\
&-\Bigg[\sum_{\substack{j\leq n\\j\neq i}}\left(\hat{v}_{(j)}-\hat{\gamma}_{(j)}\sum_{w=0}^{r_i-1}p_w\right)+\sum_{w=\overline{r}^i}^{i-1}\hat{\gamma}_{(w)}p_w\\
&\qquad\qquad\qquad\qquad\qquad\qquad+\overline{v}^i-\overline{\gamma}^i\sum_{w=0}^{\overline{r}^i-1}p_w\Bigg]
\end{split}
\end{equation}
\begin{equation}\begin{split}
&=x_i\left(v_{i}-\gamma_{i}\sum_{w=0}^{r_i-1}p_w\right)+\sum_{j\neq i}x_j\left(\hat{v}_j-\hat{\gamma}_j\sum_{w=0}^{r_j-1}p_w\right)\\
&-\Bigg[\sum_{\substack{j\leq n\\j\neq i}}\left(\hat{v}_{(j)}-\hat{\gamma}_{(j)}\sum_{w=0}^{r_i-1}p_w\right)+\sum_{w=\overline{r}^i}^{i-1}\hat{\gamma}_{(w)}p_w+\overline{\theta}^i\Bigg]
\end{split}
\end{equation}
\begin{equation}\nonumber\label{exppayoff_v_i3}\begin{split}
&=x_i\left(v_{i}-\gamma_{i}\sum_{w=0}^{r_i-1}p_w\right)+\sum_{j\neq i}x_j\left(\hat{v}_j-\hat{\gamma}_j\sum_{w=0}^{r_j-1}p_w\right)-V^{-i}\\
&=V^*-V^{-i}\geq 0.
\end{split}
\end{equation}

Here, as in Case 1 and Case 2 the last equality is due to Lemma 2. Again bidding truthfully maximizes LSE $i$'s expected payoff, and this expected payoff is nonnegative. Therefore in all cases, the mechanism is incentive compatible and individually rational in expectation for LSEs $i\in\mathcal{I}$. 

Before moving to the situation for an LSE $j\notin\mathcal{I}$, observe that in each case presented thus far, the expected payment made by LSE $i\in\mathcal{I}$ has been shown to be the difference between the expected SW enjoyed by LSEs $j\neq i$ when LSE $i$ is or is not considered in the generator's selection. In fact, omitting the expectation from the preceding calculations shows that given any $\hat{\sigma}_{-i}$ and generation realization $w$, the payment made by LSE $i$ is equal to the externality imposed on the other LSEs. This is the key to the IC and IR in expectation properties just demonstrated for LSEs $i\in\mathcal{I}$. 

Considering an LSE $j\notin\mathcal{I}$, $t^r_j=t^d_j=0$. Again, here it can be seen that since $V^{-j}=V^*$, LSE $j$ imposes zero externality, so that they too make an overall payment equal to their associated externality. Therefore the mechanism is IC and IR in expectation (see \cite{clarke1971multipart}, \cite{groves1975incentives}). 

Therefore in all cases, regardless of whether LSE $i$ is selected in $\mathcal{I}$, truthful bidding is a weakly dominant strategy in expectation which achieves nonnegative expected payoff. This shows that the mechanism achieves incentive compatibility and individual rationality in expectation. Given that all LSEs are incentivized to bid truthfully, we can assume that the generator makes its selections and deselections with the true LSE information $\sigma$, so that selection and deselection schemes (\ref{selectscheme}) and (\ref{secondstage_select}) are efficient. $\hfill\blacksquare$

\section{Conclusions}\label{sec:conclusions}

In this work, we propose a two-stage mechanism for the sale of stochastic power. While our mechanism is shown to be incentive compatible and individual rational in expectation, we also demonstrate that for specific generation scenarios it is possible to give explicit descriptions of payments for both stages. Further, the component day ahead and real time payments have natural interpretations, e.g., credit or compensation for shortfall averted or incurred, respectively. 

In the future, we will incorporate grid transmission constraints in order to make our setting more realistic. Also we will explore how to bring together existing one sided mechanisms for buying stochastic power with the one presented here in order to form two sided exchanges serving both generators and LSEs.

This work, as well as the proposed future work has the potential to facilitate wider adoption of renewables into smart grid networks. 


\bibliography{sellrandom}
\bibliographystyle{plain}

\begin{appendices}
\section{Proof of Lemma 1}

\begin{proof}
Since $\mathcal{I}$ is the optimal selection, it must be true that adding any other LSE $j\notin\mathcal{I}$ would decrease the expected SW achieved. Let $V'$ be the expected social welfare achieved when LSE $j$ is added to $\mathcal{I}$. Then
\begin{equation}0\geq V'-V^*=\hat{v}_j-\hat{\gamma}_jp_0-\sum_{w=1}^{n}\min(\hat{\gamma}_j,\hat{\gamma}_{(w)})p_w\end{equation}
The first two terms on the far right represent the value LSE $j$ would contribute, and additional cost incurred if no units are produced. The $\min$ terms in the sum express that given $w$ units are produced, whether LSE $(w)$ would still be the ``last'' LSE to receive a unit as under selection $\mathcal{I}$, or if instead LSE $j$ would receive the unit. The LSE with the larger $\hat{\gamma}$ value receives the unit. 
\end{proof}

The last inequality in Lemma 1 shows that in fact no selection should include more than $n-1$ LSEs which are not in $\mathcal{I}$, because those with rank $n-1+k$ with $k>0$ will make expected SW contribution
\begin{equation}\begin{split}\hat{v}_{(n-1+k)'}-\hat{\gamma}_{(n-1+k)'}\sum_{w=0}^{n-1+k}p_w&\leq\\
\hat{v}_{(n-1+k)'}-\hat{\gamma}_{(n-1+k)'}\sum_{w=0}^{n}p_w&\leq 0 \end{split}\end{equation}

\section{Proof of Lemma 2}
\begin{proof} 
Starting with the selection $\mathcal{I}$, since $x_i=1$ if, and only if $i\in\mathcal{I}$, the expected SW achieved by $\mathcal{I}$ is 
\begin{equation}\label{vstar}\mathbb{E}[SW(\mathcal{I},W;\hat{\sigma})]=\sum_{i=1}^n\left(\hat{v}_{(i)}-\hat{\gamma}_{(i)}\sum_{w=0}^{i-1}p_w\right)=V^*\end{equation}
Consider another selection $\tilde{\mathcal{I}}$ which simply swaps LSE $(i)$ for another LSE $j\notin\mathcal{I}$, keeping LSE $j$ in rank $i$. Note that this may be suboptimal, even amongst all selections which include $\mathcal{I}\backslash{\{(i)\}}\cup\{j\}$. The expected SW achieved by $\tilde{\mathcal{I}}$ is 
\begin{equation}\label{vprime}\sum_{\substack{k\leq n\\k\neq i}}\left(\hat{v}_{(k)}-\hat{\gamma}_{(k)}\sum_{w=0}^{k-1}p_w\right) + \hat{v}_j-\hat{\gamma}_j\sum_{w=0}^{i-1}p_w\leq V^*\end{equation}
In particular, subtracting the first sum from both sides of (\ref{vprime}) and using (\ref{vstar}) gives 
\begin{equation}\label{lemma2}\hat{v}_j-\hat{\gamma}_j\sum_{w=0}^{i-1}p_w\leq \hat{v}_{(i)}-\hat{\gamma}_{(i)}\sum_{w=0}^{i-1}p_w\end{equation}
Note that (\ref{lemma2}) holds for any $i\leq n$ and $j\notin\mathcal{I}$. 

(\ref{lemma2}) shows that an arbitrary selection $\mathcal{I}'$ with $|\mathcal{I}'|=n+m$ and $m>1$, and $\mathcal{I}\cap\mathcal{I}'=\emptyset$, can be improved by substituting LSEs $\mathcal{I}\backslash\{(i)\}$ for the $n-1$ LSEs in $\mathcal{I}'$ with the corresponding ranks. Associating with $\mathcal{I}'$ allocation $x'$ and rankings $r'$, denoted $(\cdot)'$, yields
\begin{equation}\begin{split}\mathbb{E}&[SW(\mathcal{I}',W;\hat{\sigma})]=\sum_{j=1}^{n'}\left(\hat{v}_{(j)'}-\hat{\gamma}_{(j)'}\sum_{w=0}^{j-1}p_w\right)
\end{split}\end{equation}
\begin{equation}\label{rawsum}\begin{split}
\leq \sum_{\substack{j\leq n\\j\neq i}}\left(\hat{v}_{(j)}-\hat{\gamma}_{(j)}\sum_{w=0}^{j-1}p_w\right) &+\hat{v}_{(i)'}-\hat{\gamma}_{(i)'}\sum_{w=0}^{i-1}p_w\\
&+ \sum_{j=n+1}^{n+m}\left(\hat{v}_{(j)'}-\hat{\gamma}_{(j)'}\sum_{w=0}^{j-1}p_w\right)
\end{split}\end{equation}

(\ref{rawsum}) contains contributions from LSEs from $\mathcal{I}\backslash\{(i)\}$ as well as $\mathcal{I}'$. As written, (\ref{rawsum}) assumes they are ranked in order (top to bottom rank)
\begin{equation}(1),\dots,(i-1),(i)',(i+1),\dots,(n),(n+1)',\dots,(n+m)'\end{equation}
This ordering may not agree with the optimal order corresponding to the LSEs $\hat{\gamma}$ values. Without listing the $\hat{\gamma}$ values in order, it is still possible to determine the contribution an LSE $j\in\{(n+1)',\dots,(n+m)'\}\subset\mathcal{I}'$ would make to the expected SW once the LSEs are ranked according to $\hat{\gamma}$ values via the following expression. Let $1\leq k\leq m$. Then this contribution is
\begin{equation}\label{contribution}\begin{split}\hat{v}_{(n+k)'}-\hat{\gamma}_{(n+k)'}\sum_{w=0}^{k}p_w-\sum_{w=k+1}^{k+i-1}\min(\hat{\gamma}_{(n+k)'},\hat{\gamma}_{(w-k)})p_w\\
-\sum_{w=k+i}^{k+n-1}\min(\hat{\gamma}_{(n+k)'},\hat{\gamma}_{(w-k+1)})p_w\end{split}\end{equation}
The first two terms in (\ref{contribution}) reflect the value that $\hat{v}_{(n+k)'}$ adds, along with the potential cost due to their relative ranking amongst LSEs $\{(i)',(n+1)',\dots,(n+m)'\}$. The second two sums compare LSE $(n+k)'$ with the LSEs from $\mathcal{I}\backslash\{(i)\}$ in order to determine, for each generation level $k<w\leq k+n$, which would receive a unit. 
Continuing,
\begin{equation}\begin{split}
(\ref{contribution})\leq \hat{v}_{(n+k)'}&-\hat{\gamma}_{(n+k)'}p_0-\sum_{w=1}^k\min(\hat{\gamma}_{(n+k)'},\hat{\gamma}_{(w)})p_w\\
&-\sum_{w=k+1}^{k+i-1}\min(\hat{\gamma}_{(n+k)'},\hat{\gamma}_{(w-k)})p_w\\
&-\sum_{w=k+i}^{k+n-1}\min(\hat{\gamma}_{(n+k)'},\hat{\gamma}_{(w-k+1)})p_w\\
\end{split}\end{equation}
and here using that when $k\geq 1$, $\hat{\gamma}_{(w)}<\hat{\gamma}_{(w-k+1)}<\hat{\gamma}_{(w-k)}$
\begin{equation}\begin{split}
\leq \hat{v}_{(n+k)'}&-\hat{\gamma}_{(n+k)'}p_0-\sum_{w=1}^k\min(\hat{\gamma}_{(n+k)'},\hat{\gamma}_{(w)})p_w\\
&-\sum_{w=k+1}^{k+i-1}\min(\hat{\gamma}_{(n+k)'},\hat{\gamma}_{(w)})p_w\\
&-\sum_{w=k+i}^{n}\min(\hat{\gamma}_{(n+k)'},\hat{\gamma}_{(w)})p_w\\
\end{split}\end{equation}
\begin{equation}\begin{split}
&\leq \hat{v}_{(n+k)'}-\hat{\gamma}_{(n+k)'}p_0-\sum_{w=1}^n\min(\hat{\gamma}_{(n+k)'},\hat{\gamma}_{(w)})p_w\\
&\leq 0
\end{split}
\end{equation}
where the last inequality follows from Lemma 1. Therefore when the full set of LSEs included in (\ref{rawsum}) are ranked according to $\hat{\gamma}$ values in order to maximize expected social welfare, LSEs $\{(n+1)',\dots,(n+m)'\}$ will actually make negative contributions, and therefore may be discarded. Aside from LSEs $\mathcal{I}\backslash\{(i)\}$ this leaves only $(i)'$, which makes an expected social welfare contribution of
\begin{equation}\begin{split}\hat{v}_{(i)'}-\hat{\gamma}_{(i)'}p_0-\sum_{w=1}^{i-1}\min(\hat{\gamma}_{(i)'},\hat{\gamma}_{(w)})p_w\\-\sum_{w=i}^{n-1}\min(\hat{\gamma}_{(i)'},\hat{\gamma}_{(w+1)})p_w\end{split}\end{equation}
Note that this is precisely $\theta^i_{j}$ as defined in (\ref{thetadef}) for $j=(i)'$. The optimal LSE $j\notin\mathcal{I}$ to add, then, is the one which maximizes $\theta^i_j$. Thus the optimal selection, $\mathcal{I}^{-(i)}$ which ignores LSE $(i)$ will select $\mathcal{I}\backslash\{(i)\}$ when $\overline{\theta}^i\leq 0$ and $\mathcal{I}\backslash\{(i)\}\cup\{j(i)\}$ when $\overline{\theta}^i>0$.
\end{proof}

\end{appendices}

\end{document}